\documentclass[runningheads]{llncs}

\usepackage[T1]{fontenc}

\usepackage{tabularray}
\usepackage{graphicx}
\usepackage[table,xcdraw]{xcolor}
\usepackage{hyperref}
\usepackage{siunitx}
\usepackage{afterpage}
\usepackage{algorithmic}
\usepackage{array} %
\usepackage{amsmath,amssymb,amsfonts}
\usepackage{booktabs} %
\usepackage{color}
\usepackage{graphbox}

\usepackage{mathtools}
\usepackage{multirow}
\usepackage{textcomp}
\usepackage{tikz}
\usepackage{pifont} %
\usepackage{colortbl}
\usepackage{flushend}
\usepackage[normalem]{ulem}
\useunder{\uline}{\ul}{}
\usepackage{hyperref}
\usepackage[capitalise]{cleveref} 
\usepackage{atbegshi} 
\usepackage{todonotes}
\usepackage{mfirstuc}
\usepackage{csquotes}

\usepackage{textcomp}

\begin{document}

\title{Service-Level Energy Modeling and Experimentation for Cloud-Native Microservices}
\titlerunning{Service-Level Energy Modeling for Cloud-Native Microservices}
\author{Julian Legler \and
Sebastian Werner \and
Maria C. Borges \and 
Stefan Tai
}

\authorrunning{Legler et al.}

\institute{Technische Universität Berlin, Information Systems Engineering \\
\url{https://tu.berlin/ise} \\
\email{\{jl,sw,mb,st\}@ise.tu-berlin.de}}
\maketitle              %
\begin{abstract}
Microservice architectures have become the dominant par\-adigm for cloud-native systems, offering flexibility and scalability. However, this shift has also led to increased demand for cloud resources, contributing to higher energy consumption and carbon emissions. While existing research has focused on measuring fine-grained energy usage of CPU and memory at the container level, or on system-wide assessments, these approaches often overlook the energy impact of cross-container service interactions, especially those involving network and storage for auxiliary services such as observability and system monitoring.
To address this gap, we introduce a service-level energy model that captures the distributed nature of microservice execution across containers. Our model is supported by an experimentation tool that accounts for energy consumption not just in CPU and memory, but also in network and storage components. We validate our approach through extensive experimentation with diverse experiment configurations of auxiliary services for a popular open-source cloud-native microservice application. Results show that omitting network and storage can lead to an underestimation of auxiliary service energy use by up to 63\%, highlighting the need for more comprehensive energy assessments in the design of energy-efficient microservice architectures.

\end{abstract}
\section{Introduction}\label{ch:intro}

Modern software systems are increasingly built using microservice architectures, particularly in cloud-native environments. This architectural style promotes scalability and flexibility. However, the decomposition of applications into independently deployable services can significantly increase the overall demand for computing resources. Each service may incur its own overhead in terms of computation, networking, and storage.

The International Energy Agency (IEA) projects that data-centre electricity demand will nearly double to around 945 TWh, which will be just short of 3\% of global energy consumption, by 2030, driven largely by AI and cloud workloads \cite{iea_energy_2025}. Notably, the storage energy costs alone can account for up to 33\% of a data-centre’s operational emissions \cite{mcallister_storage_emission_hot_carbon_2024}.  

While the environmental footprint of cloud-native systems has drawn growing concern, most energy-efficiency efforts to date have focused on either measuring low-level resource usage (e.g., CPU and memory within containers) or coarse-grained system-wide measurements. This leaves a critical gap: service-level energy accounting in distributed deployments, where a single service spans multiple containers and often multiple hosts. Existing approaches also overlook cross-container interactions, especially those involving auxiliary services (e.g., for observability, logging, authentication, and other) that rely heavily on network and storage subsystems. These components can represent a substantial portion of a system's energy use, especially in highly distributed service-oriented architectures.

To address this gap, we propose a service-level energy model tailored for cloud-native microservice architectures. 
Our approach explicitly includes modeling of distributed service executions by means of inter-container communication, thereby accounting for energy consumption of network and storage for all services. 
Our service-level energy model is implemented via a tool -- GOXN -- that enables detailed energy measurements for individual microservices (the primary services) and for auxiliary system services in support of the primary services and their interactions. 
The tool is a significant extension of the previously published generic Observability eXperiment eNgine (OXN)~\cite{borges_informed_observability_decisions_2024}. It extends OXN by adding full Kubernetes support, adding energy measurement support, adding multiple cloud-native treatments, and heavily extending OXN's portability.

We experimentally validate our model using a widely adopted open-source microservice application. 
By comparing alternative designs for auxiliary services, we demonstrate the impact of including network and storage components in energy assessments that so far have been underestimated.

In this paper, we thus address the two research questions of:
\begin{description}
    \item[R1]\textit{How can we model and measure the energy consumption of cloud-native microservices at the service level, including auxiliary services, and by capturing distributed interactions across CPU, memory, network, and storage?}
\vspace{1em}
\item[R2] \textit{What tool support is needed to measure and compare the energy efficiency of alternative microservice designs, especially considering auxiliary services?}
\end{description}
\noindent{}Toward this end, we present the following main contributions:
\begin{description}
    \item[C1:] We propose a fine-grained, service-level energy consumption model for micro\-service-based systems. The model accounts for the operational energy use of CPU, memory, network, and storage, enabling a more precise assessment of energy efficiency across distributed services.
    \item[C2:] We provide an open-source implementation of our model by extending the Observability eXperiment eNgine (OXN), enhancing it with energy measurement capabilities across multiple resource types.
    \item[C3:] We conduct an experimental evaluation using a representative cloud-native demo application, demonstrating the tool's ability to compare alternative service designs and uncover energy inefficiencies, particularly in auxiliary services.
\end{description}

In the following Section \ref{ch:bgrw}, we give a short overview of the state-of-the-art in operational energy measurement of microservice systems. %
Section \ref{ch:model} introduces our first contribution, a service-level energy model for cloud-native microservice systems. In Section \ref{ch:design}, we present our second contribution, the service-level energy measurement system GOXN, an energy experimentation framework that can measure the consumed energy on the service level, incorporating CPU, memory, storage, and network. This is followed by Section \ref{ch:eval}, where we describe our experimental approach to evaluation. In Section \ref{ch:result}, we present and discuss our experiment results and conclude in Section \ref{ch:conclusion}.

\section{Background and Related Work}\label{ch:bgrw}
In order to model and measure the energy costs of microservices, we first differentiate the types of services that typically exist in these architectures.
Specifically, we differentiate between \textbf{Primary Services}, i.e., business-fulfilling services, and \textbf{Auxiliary Services}, i.e.,  services that support primary services but do not directly contribute to the business functionality.

While primary services are often developed by individual teams and may already measure and consider energy efficiency~\cite{werner_clue_2025}, auxiliary services are often shared by all teams in an organization to address non-functional design objectives, such as maintainability through a tracing service. Additionally, they are often not designed from scratch but instead configured commodity systems \cite{li_survey_microservice_tracing_2022}. 

Some studies have already shown that auxiliary services, such as tracing systems, can introduce significant overhead in terms of energy consumption~\cite{dinga_energyoverheadobservability_2023}.
However, these studies focused on the compute-based overhead of auxiliary services and did not consider the energy overhead of network and storage.
Others, such as Wang et. al.~\cite{wang_green_service_composition_2021}, are investigating energy-efficient service composition, hence, the interaction between primary and auxiliary services.
They identify the network as an energy consumption source and try to do a more energy-efficient service composition by ensuring services are closer together or on the same node.
Zhao et al. \cite{zhao_granularity_energy_impact_2025} recently showed, in a controlled experiment on two cloud-native microservices systems, that finer microservice granularity can raise energy use by up to 13 \% and response time by 14 \%, probably also related to the extra inter-service calls required.  
Complementing this, Xiao et al. \cite{xiao_microservie_pattern_energy_2025} evaluated six common microservice tactics and patterns and found that communication-oriented techniques reduce energy most effectively but often trade maintainability or performance for these improvements.  
Both studies focus on compute energy consumption relying either on Scaphandre or Powerstat as measurement tools.
Therefore, network and storage overhead that their techniques introduce remains unaccounted for.  
Our work fills this gap by quantifying exactly those hidden energy costs and showing that, for auxiliary services, network and storage can dominate compute-level energy by up to 70 \% (see~\Cref{ch:result}).

We argue that, specifically, auxiliary services often use more internal network and storage resources, and therefore, their energy use is often underestimated when only observing direct CPU and memory consumption. Therefore, we must not only consider the service-based design decisions but also the broader composition of services and their role.
In \Cref{ch:result}, we quantify the share of auxiliary-service energy in a microservice application to showcase this underestimation.
Nevertheless, in general, it is important to measure the energy consumption not only on a compute-based level but also on a network and storage level to get a complete picture of the energy consumption of microservices to make informed design decisions.

When it comes to measuring the energy consumption of microservices, especially in containerized environments, practitioners and researchers have to rely on compute-based energy monitoring tools, such as Scaphandre and Kepler~\cite{amaral_kepler_2023}.
Centofanti et al.~\cite{centofanti_analysis_open_source_power_measurement_tools} compared these tools to traditional methods, highlighting their utility in Kubernetes settings due to their support for cloud-native deployments and container-level energy metrics.
While these tools are not perfect~\cite{pijnacker_container_level_observability_2025}, they are already used in practice to measure the energy consumption of microservices~\cite{werner_clue_2025}.
However, because these tools only provide information on CPU and memory consumption, prior work that utilizes them tends to focus solely on these aspects, without considering network or storage, which can lead to potential blind spots in their assessments.

Unfortunately, no established tooling for network and storage exists, so researchers rely on energy-intensity (EI) factors to estimate the energy consumption of services based on observable measurements, such as traffic volume or storage operations.
In the case of network consumption, network energy-intensity provides us with a kWh/GB estimation that accounts for all the energy consumed by the network infrastructure, including routers, switches, and other devices, to transmit data across the network.
These estimates range from around $0.06\text{ kWh/GB}~$\cite{aslan_kwh_per_gb_2018} in 2015 to $0.02 \text{ kWh/GB}$~\cite{study_on_energy_intensities_2021} in 2020, suggesting a gain in energy efficiency of network infrastructure over the years.
In the case of storage consumption, storage energy-intensity factors provide an estimation of kWh/GB per hour, accounting for the energy consumed by storing data and maintaining it for a specified period, such as one year.
Here, for storage, Al Kez et al. \cite{al_kez_exploring_2022} found that maintaining one GB of data for a year uses $0.0046 \text{ kWh}$.

However, so far, there are no approaches that combine these measurements to evaluate service design decisions in microservice architectures, for both primary and auxiliary services.
Hence, we derive a more inclusive model for estimating the energy consumption of microservices by combining compute energy measurements with network and storage energy-intensity factors, as presented in \Cref{ch:model}, and use it to close this gap.

\section{Service Level Energy Model}\label{ch:model} %

We now present a service-level energy model for cloud-native microservice architectures to answer \textbf{R1}. As motivated above, our model includes critical energy-consuming hardware components such as CPU and memory, but also network and storage. We map these to conceptual service components using Kubernetes terminology of containers, pods, nodes, and services as industry-wide accepted microservice architecture building blocks. 
The model expands the Component-Level Additive Model as discussed by Dayarathna et al.~\cite{Dayarathna_energy_consumption_modeling_2016}. 

\begin{figure}
    \centering
    \includegraphics[width=0.85\columnwidth]{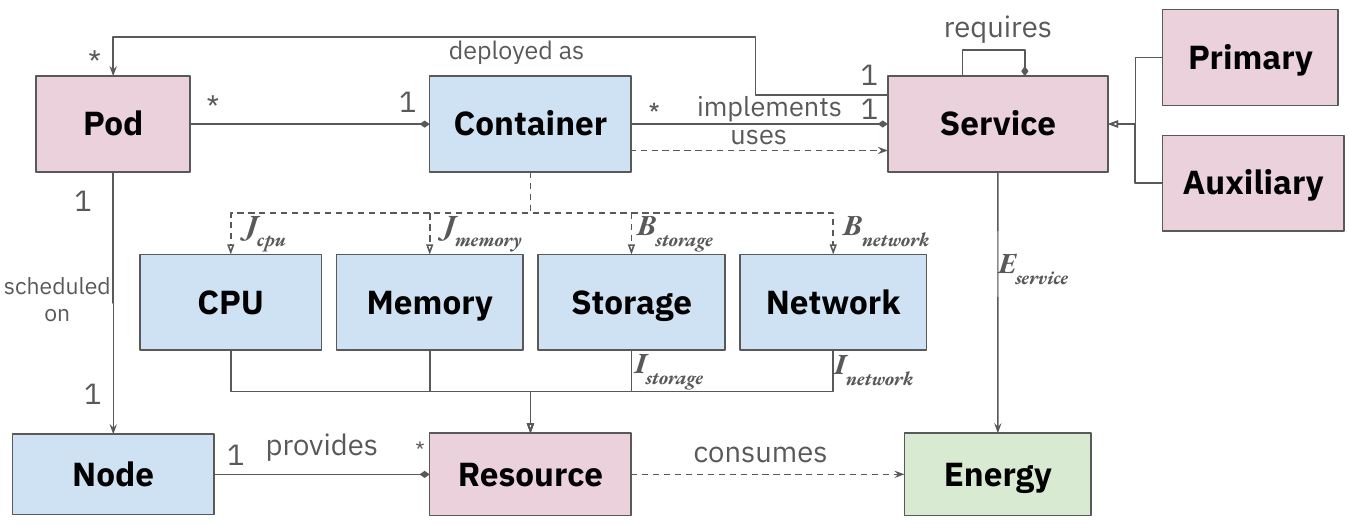}
    \caption{Conceptual Microservice Architecture, oriented on Kubernetes, highlighting the relevant measurement points (blue) and inter-container and inter-service interactions.}\label{fig:conceptional_model}
\end{figure}

\subsection{Service-Level Energy Consumption Model}

In the context of making informed design decisions in microservice architectures, it is crucial to measure energy consumption not only at coarse-grained levels (e.g., nodes or data centers) and fine-grained levels (e.g., containers or processes) but also primarily at the service level, where such decisions have the greatest impact. Existing tooling provides the coarse and fine levels, yet falls short of offering an easy-to-use approach for service-level energy consumption.

A service architecture may consist of multiple pods distributed across several nodes. Therefore, in order to assess whether a particular design decision is justified, it is necessary to measure the energy consumption at the service level, rather than limiting the assessment to individual pods or containers. Furthermore, certain services may rely on auxiliary components, such as observability services or service mesh frameworks, which should also be considered in a comprehensive evaluation.

Thus, we define, as depicted in Figure~\ref{fig:conceptional_model}, a service as a collection of containers that are deployed to achieve a specific functionality, e.g., a primary service to handle business functions such as ordering, or an auxiliary service to support primary services such as a tracing service.
It is important to distinguish between the way a service is deployed, represented by a pod, and the service to which a container belongs. For instance, a sidecar container used for collecting metrics may be associated both with the primary service of the pod and with an auxiliary monitoring service.

From this conceptual model, we define the total energy consumption of a service (\(E_{\text{service}}\)) as the sum of the energy consumption across all its containers:

\begin{align}
E_{\text{service}} &= \sum_{j \in P} E_{\text{container}, j}
\label{equation:service-energy}
\end{align}
\noindent
where $P$ is the set of all service containers, and $E_{\text{container}, j}$ is energy of \( j \)-th container
Each container's energy consumption, in turn, is composed of several component energies:
\begin{equation}\label{equation:pod-component-energy}
E_{\text{container}, j} = E_{\text{cpu}, j} + E_{\text{memory}, j} + E_{\text{network}, j} + E_{\text{storage}, j}
\end{equation}
where \( E_{\text{cpu}, j} \) is the energy consumption due to CPU usage, \( E_{\text{memory}, j} \) is the energy consumption due to memory usage, \( E_{\text{network}, j} \) is the energy consumption due to network usage and \( E_{\text{storage}, j} \) is the energy consumption related to storage of the container \( j \), respectively.

For each usage component, it is necessary to define the method by which energy consumption is calculated. This requires distinguishing between directly accountable components, such as CPU and memory, and indirect components, such as network and storage, which rely on energy intensity models to convert usage data into energy consumption estimates~\cite{aslan_kwh_per_gb_2018}.

\paragraph{Indirect Accountable Components,} specifically, storage and network are defined as follows:
\noindent
\begin{equation}\label{equation:service-network-energy}
    E_{\text{network}, j} = \frac{B_{\text{network}, j}}{10^9} \cdot I_{\text{network}} 
\end{equation}
\noindent
where \( B_{\text{network}, j} \) is the total bytes transmitted over the network by the \( j \)-th container and \( I_{\text{network}} \) is the energy intensity \( (kWh/GB) \) for network data. 
Accordingly, we define the storage energy consumption as follows:
\begin{equation}\label{equation:service-storage-energy}
    E_{\text{storage}, j} = \frac{B_{\text{storage}, j}}{10^9} \cdot I_{\text{storage}}
\end{equation}
with \( B_{\text{storage}, j} \) is the total bytes used for storage by the \( j \)-th container and \( I_{\text{storage}} \) is the energy intensity \( (kWh/GB) \) for storage data.

\paragraph{Directly Accountable Components,} such as CPU and memory usage, are defined as follows:
\begin{align}
    E_{\text{cpu}, j} = \frac{J_{\text{cpu}, j}}{3600000} & \text{, }
    E_{\text{memory}, j} = \frac{J_{\text{memory}, j}}{3600000},
    E_{\text{compute}, j} = E_{\text{cpu}, j} + E_{\text{memory}, j},
\end{align}
where \( J_{\text{cpu}, j} \) is the joules consumed by the CPU and \( J_{\text{memory}, j} \) is the joules consumed by the memory.

\paragraph{\textbf{Total Energy Consumption}}
can then be calculate the in \textbf{$kWh$} by summing the energy consumed by all services $S$, as follows:
\begin{equation}\label{equation:total-system-energy}
\mbox{\large\(
    E_{\text{total}}(S) = \sum_{k}^{S} E_{\text{service}, k} + \frac{1}{|S|}E_{\text{system}}
\)}
\end{equation}

where:
\begin{itemize}
    \item \( S \) is the set of all services in the microservice architecture, including all auxiliary services such as observability systems.
    \item \( E_{\text{service}, k} \) is the energy consumed by the \( k \)-th service.
    \item \( E_{\text{system}} \) is the energy consumption by system processes not attributable to any specific service.
\end{itemize}

Note that our model differs from the Kubernetes deployment model by assuming that functionality is also bundled in composed services, e.g., we can introduce a virtual observability service that composes a logging, tracing, and metrics collector service, while in the Kubernetes model, these would all be separate entities. 
Hence, we can, for example, calculate the \textbf{ total energy of the auxiliary service} of all observability services as we do in \cref{ch:eval}, by selecting the virtual composed services in $S_{auxiliary} \subset S$, which is defined as any service in S that contributes to the virtual observability service and applying $ E_{\text{total}}(S_{auxiliary})$ as described.

\subsection{Deliberate Constraints}

The primary objective of the model is to enable mapping of the components of a microservice architecture to the energy consumption of the underlying hardware, specifically including network and storage.
However, we must acknowledge that not all of the required measurements can be easily obtained.

For example, the energy consumption of the network measured through the total transmitted bytes may include double-accounted traffic, when an internal service is communicating with another internal service; thus, the energy consumption of the network may be overestimated.
Here, techniques such as network packet tagging can be used to differentiate and account for internal and external traffic. 
Then again, not accounting for the network-related consumption due to complex measurements or the risk of double accounting can lead to designs that shift consumption to unmeasured parts of the hardware.

Additionally, the energy intensity values \( I_{\text{network}} \text{ and } I_{\text{storage}}\) significantly depend on the specific hardware and network configuration of the data center, and can vary widely across different environments~\cite{hossfeld_energy_intensity_used_wrong_2024}.
Here, several studies exist that propose specific intensities that can be used as a starting point, such as \cite{aslan_kwh_per_gb_2018,study_on_energy_intensities_2021}. 
Naturally, data center and cloud providers could calculate and publish the intensities for their systems to make this model more accurate.
Nevertheless, using the existing intensity values for metastudies already provides a reasonable estimate that helps to inform design decisions, even if the real consumption might be slightly over- or underestimated.

\section{Service-Level Energy Measurement System}\label{ch:design}

This section introduces the Green Observability eXperiment eNgine (GOXN), our holistic, fine-grained open-source service-level energy measurement tool capable of measuring and attributing the operational energy consumption of CPU, memory, storage, and network for microservice applications, answering \textbf{R2}\footnote{Available at: \url{https://github.com/JulianLegler/goxn}}. 

We based GOXN on the existing open-source experimentation framework OXN~\cite{borges_informed_observability_decisions_2024}. We did significant additions and extensions to OXN to run in cloud-native environments and enable service-level energy consumption measurements.

\subsection{Requirements}
The primary requirements for GOXN include the ability to measure energy consumption at the service level, to execute and monitor cloud-native microservice applications automatically, and to support reproducible experimentation. The core objective of the tool is to enable practitioners to define and evaluate changes to their microservice architecture, service configurations, or auxiliary service integrations.

GOXN utilizes established cloud-native tools for automatic deployment, such as Kubernetes-Helm, as well as industry standards for collecting measurement data, including Prometheus, Kepler, and cAdvisor. This enables easy integration into existing staging environments or the use of a dedicated cluster for experiments.
To gather the needed information for our Service-Level Energy Model, introduced in \cref{ch:model}, the system needs to measure disk operations including reading and writing data ($B_{storage, j}$), sending and receiving data via network ($B_{network, j}$), energy usage of the CPU ($J_{cpu, j}$) and memory ($J_{memory, j}$).
Additionally, GOXN needs to expose an easy and extensible mechanism to define and run treatments, e.g., changes to the service composition, configuration changes, or replacement of services.  
Moreover, the tool should provide its own load generator that is configurable to induce realistic traffic to the system when measuring consumption and to automatically collect the results at the end of the experiment to produce an easy-to-use report for practitioners performing these experiments.
Lastly, the tool should allow for estimating the consumption of the service during an analytical period, e.g., how much it would consume in 30 days.

\begin{figure}
    \centering
    \includegraphics[width=0.95\linewidth]{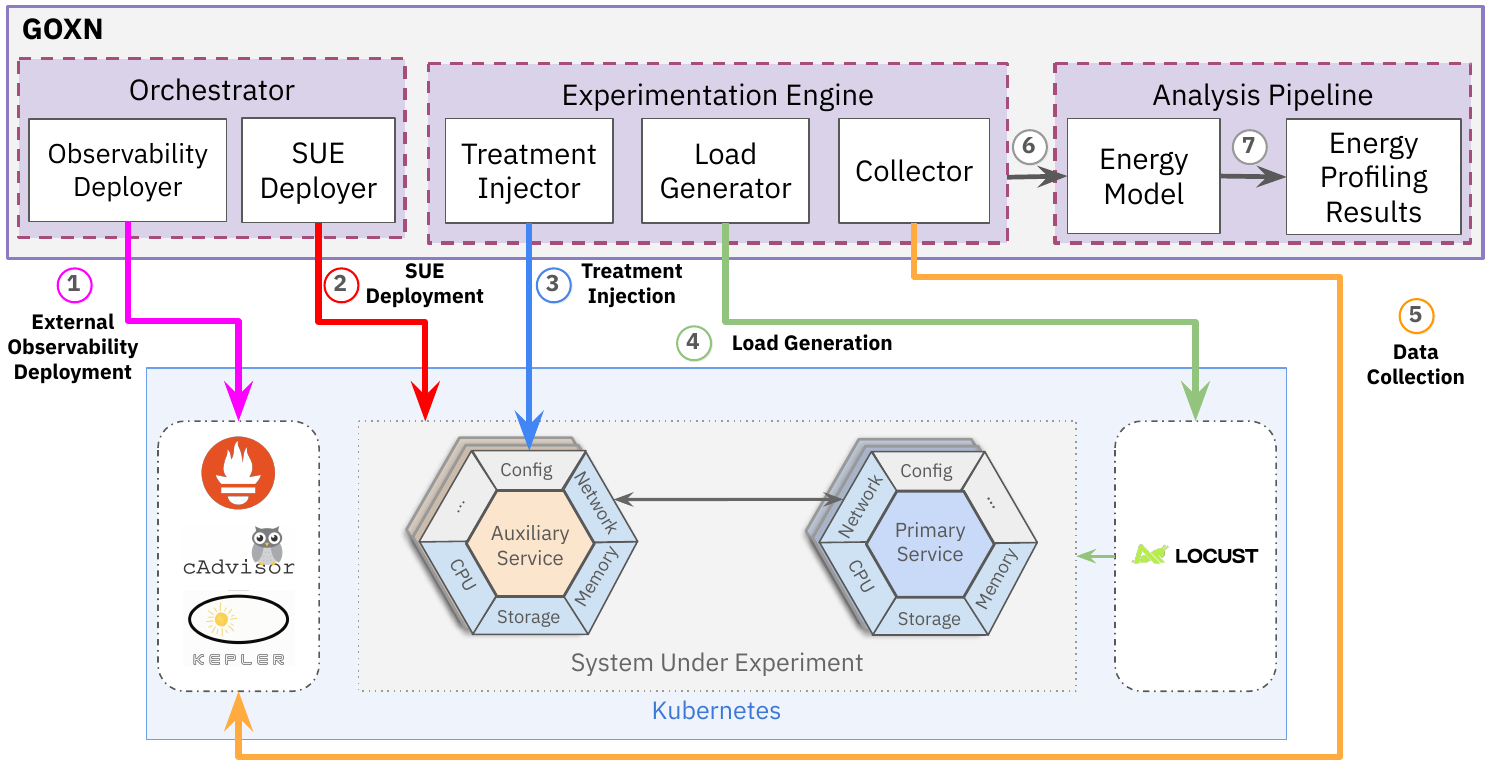}
    \caption{Architecture overview of GOXN, including a part of the SUE used in \cref{ch:eval}. Showing the different logical units of GOXN and important environment interactions.}
    \label{fig:architecture}
\end{figure}

\subsection{Architecture}

Figure \ref{fig:architecture} provides an overview of GOXN’s architecture and its integration of the System Under Experimentation (SUE) evaluated in \cref{ch:eval}. One of the primary differences from OXN is architectural. Whereas OXN was limited to local Docker deployments, GOXN introduces Kubernetes support, enabling experiments in distributed, cloud-native settings.

GOXN has three main components: (i) an experiment orchestrator -- in charge of deploying the SUE and required observability tools to capture the SUEs consumption, (ii) an experimentation engine -- capable of applying treatments to the SUE, generating load and collecting relevant measurement data during experimentation and (iii) an analysis pipeline -- used to extract and evaluate the collected measurements into an easy to handle format to evaluate the energy costs of services and treatments applied to these services.

To run an experiment, GOXN always performs the following steps:
\begin{enumerate}
    \item \textbf{External Observability Deployment:} Deployment of required observability tools into Kubernetes, i.e., a Prometheus stack, Kepler, cAdvisor.
    \item \textbf{System Under Experimentation Deployment}: Deployment of the SUE, e.g., a microservice application, that we want to evaluate. 
    Here, the user needs to supply a Helm chart.
    \item \textbf{Treatment Injection:} Once the SUE and observability stack are running, GOXN applies a defined treatment, e.g., modifying configurations. For this, a user must specify each treatment in an \textit{experimentation} config.
    \item \textbf{Load Generation:} The same \textit{experimentation} config also defines the workload profile that is used to create representative stress to the SUE, e.g., a locust profile. Hence, once the treatments are applied, GOXN starts this load generation and, in parallel, starts to collect all relevant measurements.
    \item \textbf{Data Collection:} Once the load phase is finished, GOXN queries the deployed observability stack (step 1) and stores the observed data in an HDF5 file for analysis before cleaning up the SUE and External Observability Deployment.
    \item \textbf{Energy Model Calculation:} Using the collected measurements, GOXN then calculates the energy cost for each service and stores it in the CSV file.
    \item \textbf{Analysis:} Lastly, GOXN provides Jupiter Notebook templates that can load and analyze the performed experiments. Hence, as a last optional step, the user can review and slice the collected measurements, e.g., review only the energy consumption of services directly affected by treatments.
\end{enumerate}

In a typical use case, GOXN performs these steps multiple times for each treatment, as well as for at least a baseline and a treatment, to facilitate a comparative analysis of changes to the SUE. 
For this reason, GOXN writes all results in the same HDF5 file, including the used experiment configuration that can then be analyzed together using the Jupiter Notebooks.

\subsection{Measurement Attribution Strategy}

Existing tools do not directly provide service-level energy measurements. Following our energy model, we can collect container-level energy measurements to calculate the wanted service-level metrics. The desired service-level energy components are
$E_{\mathrm{storage}}$, $E_{\mathrm{network}}$, $E_{\mathrm{cpu}}$, and $E_{\mathrm{memory}}$.
Kepler exposes the container-level metrics $E_{\mathrm{cpu}, \mathrm{j}}$ and $E_{\mathrm{memory}, \mathrm{j}}$ as
Prometheus-scraped metrics.

To measure the service-level $E_{\mathrm{storage}}$ and $E_{\mathrm{network}}$, GOXN combines the container-level proxy
metrics $B_{\mathrm{storage}, \mathrm{j}}$ and $B_{\mathrm{network}, \mathrm{j}}$ with energy intensities
$I_{\mathrm{storage}}$ and $I_{\mathrm{network}}$. cAdvisor provides container-level
metrics for bytes read from and written to storage ($B_{\mathrm{storage}, \mathrm{j}}$) and
bytes sent and received over the network ($B_{\mathrm{network}, \mathrm{j}}$), which are ingested
by Prometheus.

In addition, the GOXN records the sizes of all persistent volumes used
during an experiment and stores these values. For services that use a persistent
volume, this size is used as $B_{\mathrm{storage}}$.

The last missing values, $I_{\mathrm{storage}}$ and $I_{\mathrm{network}}$, denote the
energy intensity of storing 1\,GB of data on disk and of transferring 1\,GB of data
over a network, respectively. As discussed in \cref{ch:bgrw}, reported values for
$I_{\mathrm{network}}$ range from 0.06\,kWh/GB in 2015 \cite{aslan_kwh_per_gb_2018}
to 0.02\,kWh/GB in 2020 \cite{study_on_energy_intensities_2021}, with historical
evidence that network intensity halves approximately every two years
\cite{aslan_kwh_per_gb_2018}. Extrapolating to 2025 yields
$I_{\mathrm{network}} = 0.06/2^5 = 0.001875\,\mathrm{kWh/GB}$.

For storage, we adopt the estimate of 0.0046\,kWh to store 1\,GB for one year
\cite{al_kez_exploring_2022}. In our evaluation (\cref{ch:eval}) we focus on
observability data and assume a retention time of 30 days
\cite{karumuri_observability_data_management_retention_time_2021}, which corresponds
to $I_{\mathrm{storage}} = 0.0046 \times \frac{30}{365} \,\mathrm{kWh/GB}
\approx 0.000378\,\mathrm{kWh/GB} = 0.38\,\mathrm{Wh/GB}$ for 30 days.

Due to GOXN's extensibility, all Prometheus queries used to collect measurements, as well as
the intensity parameters, can be configured to match a user’s deployment environment.

\section{Evaluation}\label{ch:eval}
In this section, we evaluate our model and tool, aiming to demonstrate the importance of considering network and storage when calculating service-level energy consumption, particularly for auxiliary services.
Towards that aim, we designed and implemented six experiments that evaluate energy consumption implications of changes to the observability stack of a cloud-native microservice application.
We specifically selected the observability stack to replicate and confirm insights from related work by Dinga et al. \cite{dinga_energyoverheadobservability_2023}.

\subsection{Experimental Design}

\begin{figure}
\centering
\begin{minipage}{0.50\textwidth}
\centering
    \includegraphics[width=\textwidth]{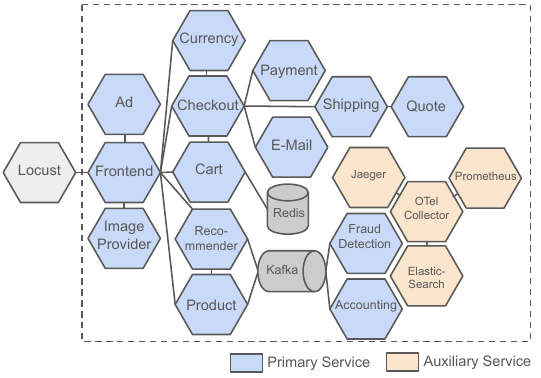}
    \label{fig:OT_arch}
\end{minipage}%
\hfill
\begin{minipage}{0.50\textwidth}
\centering
\resizebox{\textwidth}{!}{%
\begin{tabular}{@{}lccc@{}}
\toprule
Experiment & \begin{tabular}[c]{@{}c@{}}Prometheus \\ Scrape Interval\end{tabular} & \begin{tabular}[c]{@{}c@{}}Jaeger\\ Trace Sampling\\ Rate\end{tabular} & \begin{tabular}[c]{@{}c@{}}Istio\\ in Sidecar \\ deployment \end{tabular} \\ \midrule
Baseline       & 60s & 1\%  & no  \\
Monitoring Medium & 30s  & 1\%  & no  \\
Monitoring High   & 5s  & 1\%  & no  \\
Tracing Low    & 60s & 5\%  & no  \\
Tracing Medium & 60s & 10\% & no  \\
Tracing High   & 60s & 50\% & no  \\
Service Mesh   & 60s & 1\%  & yes \\ \bottomrule
\end{tabular}%
}
\end{minipage}
\caption{The SUE architecture includes 14 primary (blue) and 4 auxiliary services. The table lists experiment configurations, each run with only one controlled system change.}\label{tab:experiments}
\end{figure}

As the SUE, we selected the OpenTelemetry Astronomy Shop demo\footnote{\url{https://github.com/open-telemetry/opentelemetry-demo}}, as it focuses explicitly on observability systems and therefore offers a wide array of treatment options we can apply to auxiliary services.
Overall, we selected six treatments that modified it, as shown in \cref{tab:experiments}.

For each experiment, GOXN deploys itself and the SUE completely fresh to mitigate any unwanted behavior. 
Each individual experiment runs for 70 minutes. To avoid measuring startup or shutdown behavior, the first seven minutes and the final three minutes are excluded from the experimentation dataset. 
Leading to 60 minutes of experiment runtime. During this time, an average of 2.7 million requests are sent against the SUT using a decentralized locust deployment.
We repeated each experiment three times, yielding a result dataset including about 370 thousand measurements with a total experiment runtime of slightly over 24 hours.
The load generator can be configured via GOXN, but its configuration is the same for all experiments as a synthetic open-loop workload to enable comparability between the results. 
We tuned the workload to achieve an average request response time of below 100 ms, ensuring the system is not experiencing resource depletion or other overload symptoms. This is necessary because the experiments aim to measure the SUE under a reasonable load and are not intended to stress-test it.

For the six experiments (\cref{tab:experiments}), we implemented three treatments: Firstly, \textbf{trace sampling rate} --  the \texttt{probabilistic\_head\_sampling\_rate} treatment alters the sampling rate used by the \texttt{otel-collector} for the trace sampling. This treatment alters the sampling\_percentage in the corresponding Kubernetes ConfigMap. By default, this is set to 1\%. Here, we tested three scenarios: Low (5\%), Medium (10\%) and High (50\%) sampling rates, which should mainly affect Jaeger.
Secondly, \textbf{monitoring scrape interval} -- the interval in which Prometheus is scraping its targets, e.g., the metric endpoints of all services. The treatment \texttt{kubernetes\_prometheus\_interval} provides functionality to alter this scrape interval for a given Prometheus service. The treatment modifies the ConfigMap containing the Prometheus scrape config and then forces a restart of the service to apply the new configuration. 
Lastly, \textbf{service mesh} -- the service mesh treatment enables Istio as a service mesh into the SUE. 
The treatment deploys Istio in “Sidecar Mode” using the default configuration, following the official Istio documentation for helm charts\footnote{Istio / Install with Helm - \url{https://istio.io/latest/docs/setup/install/helm/}}. This can be used to autoinstrument all services with additional tracing points.

\subsection{Experiment Enviroment}
Experiments are conducted on a bare-metal server with an Intel(R) Xeon(R) E-2176G CPU @ 3.70GHz processor and 64 GB of DDR4 RAM, running Debian 6.1.119-1, MicroK8s v1.32.3 revision 8148, and Helm v3.18.2. Kepler is deployed using the official Helm chart version 0.6.0, which deploys Kepler release-0.8.0. As the SUE, the OpenTelemetry Demo is deployed using the official Helm chart version 0.36.4.
GOXN deploys its observability stack outside of the SUE, using the kube-prometheus-stack Helm Chart version 62.5.1.

All variables required for our Service-Level Energy Model are collected thr\-ough our experimentation system. For the final results, we aligned all measurements to Wh. We repeated the entire experiment three times to ensure reproducibility and stability.

\subsection{Experiment Artifacts}
GOXN, as an open-source project, is designed with reproducibility in mind (\cref{ch:design}).
Moreover, we provide a replication package\footnote{\url{https://github.com/JulianLegler/goxn-replication-package}}, which contains all the configuration files for conducting the presented experiments, including the Jupyter Notebooks that are used to post-process the data and all transformations done to generate the tables and plots that contain measurement data. The main readme file of the replication package includes step-by-step instructions on how to reproduce the results.

\section{Results and Discussion}\label{ch:result}
\begin{figure}[h] %
    \centering
    \includegraphics[width=0.96\linewidth]{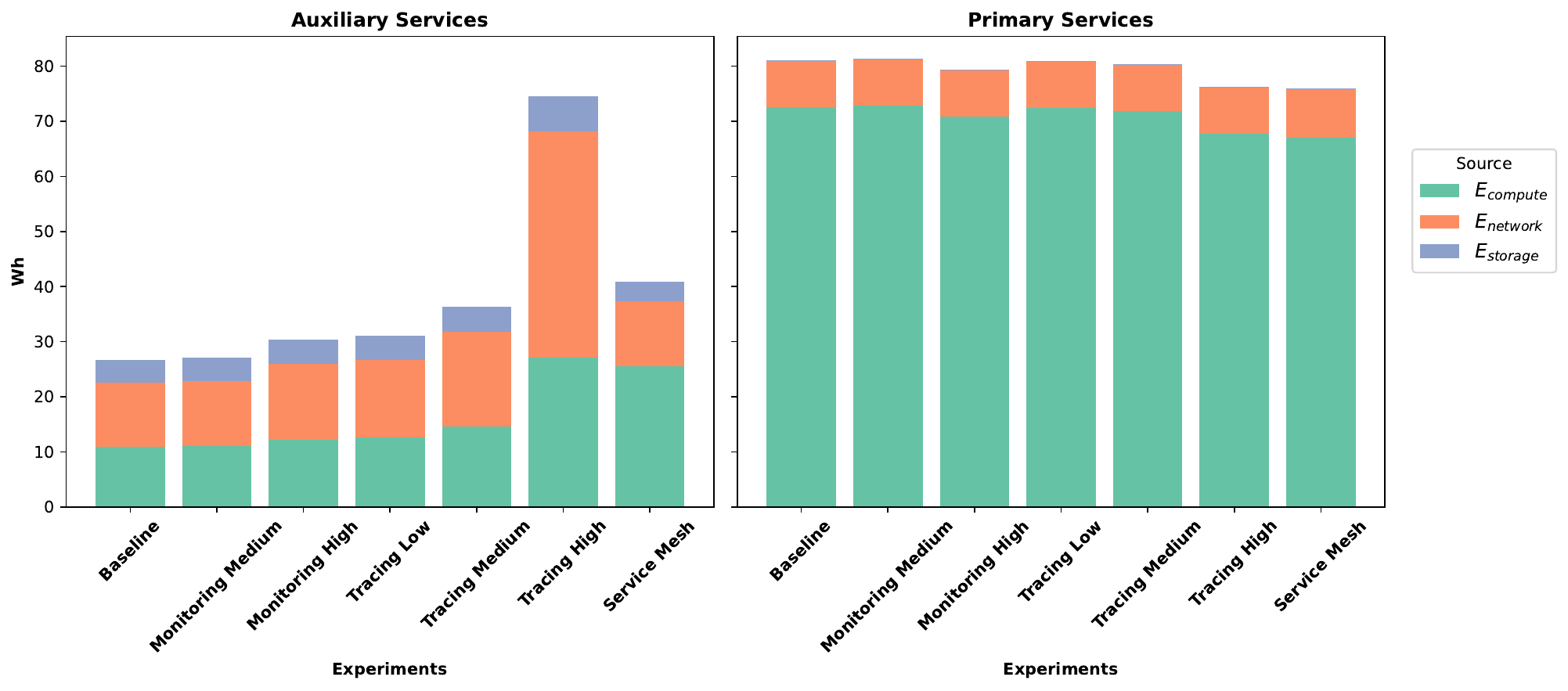}
    \caption{Ignoring the network and storage costs, as often done in the past, can result in neglecting nearly 63\% of the total energy consumption of auxiliary services. Auxiliary services are especially network heavy, and past estimations have therefore underestimated the impact of design decisions and changes.}
    \label{fig:headline_results}
\end{figure}

In this section, we analyze the energy consumption across the six experiments we defined in \cref{ch:eval}. 
We can demonstrate that neglecting network and storage can lead to a significant underestimation of energy costs that we can pinpoint to specific containers that are affected by our \texttt{Tracing High} treatment.
Showing that omitting network and storage can lead to an underestimation of auxiliary service energy use by up to 63\%.
The detailed results are presented in \cref{fig:headline_results} and \cref{tab:energy_consumption_per_service_experiment_grouped}.

\subsection{Results}
Our experiments demonstrate that design decisions, specifically for tracing systems, can increase total energy consumption by up to 47\% compared to the baseline. 
The increase is almost exclusively caused by auxiliary services, particularly by their storage and network usage. 
Across all six instrumentation scenarios, total operational energy ranged from 93.58 ± 0.30 (mean ± 95\% CI)(\texttt{Baseline}) to 137.62 ± 0.32 (\texttt{Tracing-High}). 
No significant variation is derived from primary services; auxiliary services are responsible for the vast majority of the energy increase. 
The most significant single contributor is Elasticsearch under \texttt{Tracing-High} (44.55 ± 0.30 Wh), followed by the frontend service ($\sim$ 35 Wh consistently across scenarios).

\subsubsection{Trace Sampling Rate Treatments} 
Raising the sampling rate from \SI{1}{\percent} (\texttt{Base\-line}) to \SI{50}{\percent} (\texttt{Tracing High}) multiplies the number of sampled traces by \num{49.4} (from \num{1.7e5} to \num{8.4e6}), which is expected and demonstrates the correctness of the used treatment.  
This single configuration change propagates through the entire observability stack and drives a disproportionate rise in auxiliary energy.
We see an increase from 20.1 Wh to 49.45 Wh (+146\%) in network energy, with Elasticsearch network energy consumption alone contributing 29.31 $\pm$ 0.37 Wh (before 0.6 Wh), which is a relative change of 4885\% compared to Elasticsearchs \texttt{Baseline} network energy consumption.

During the 60-minute experiment, Elasticsearch writes 66 GB yet stores only \SI{6.2}{GB}, revealing heavy write amplification; extrapolated to the 30-day retention window, constant load (500 virtual users) implies 4.5 TB of storage and 11.2 TB of inbound traffic.
In comparison, Dinga et al.~\cite{dinga_energyoverheadobservability_2023} measured an \SI{11}{\percent} system-wide energy increase for Zipkin at a comparable \SI{50}{\percent} sampling rate, whereas we record \SI{47.2}{\percent}.  
Our setting sustains \num{500} virtual users versus their \numrange{10}{40} at comparable hardware.
We would need to normalize the results by useful work, to, e.g., express the energy per trace (Wh/trace) to get comparable results to their findings. However, our work confirms the trend they identified previously.

\subsubsection{Monitoring Scrape Interval Treatments}
The \texttt{Monitoring High} scenario sees an increase from 93.58 ± 0.30 Wh to 95.71 ± 0.12 Wh (+2.4\%). 
Collecting metrics every 5 seconds instead of 60 seconds results in 86,500,000 more metrics, which are eleven times more than in \texttt{Baseline} (8,500,000). 
Prometheus receives 1.3 GB of network traffic, and storing these metrics uses 680 MB on the disk, compared to 60 MB during \texttt{Baseline}. 
This resulted in Prometheus consuming +673\% energy compared to \texttt{Baseline}, which is distributed over $E_{compute}$, $E_{network}$, and $E_{storage}$. 
While we expected to also see an overall increase in energy consumption across all services due to an increased response to metrics polling requests, we did not observe any significant rise in energy consumption in one of the primary services.
Overall, in our SUE, the impact of frequent monitoring does not appear to be significant.
This underscores the findings of Dinga et al. \cite{dinga_energyoverheadobservability_2023}, which indicate that the impact of tracing on a system's energy consumption is greater than that of monitoring. 

\subsubsection{Service Mesh Treatment}
Deploying Istio increases the energy consumption by +10.8\%. 
This is mainly due to a +133.3\% increase in compute energy overhead introduced by the newly deployed sidecar containers that are attached to every Kubernetes pod, hence, almost doubling the number of running containers in the cluster. 
If the compute energy consumption is somewhat linearly correlated with the CPU utilization, this seems to be a confirmed expected overhead of service mesh \cite{zhu_dissecting_service_mesh_2023}.
Additionally, we observe a 2\% increase in overall network energy consumption, which is most likely due to communication overhead introduced by Istio. 

\begin{table}
\centering
\caption{Energy consumption (Wh) by experiment and service; highlighting Elasticsearch (Elastic.), Jaeger, OpenTelemetry Collector (OTel), Prometheus (Prom.). Values $< 0.1$ Wh omitted.
}\label{tab:energy_consumption_per_service_experiment_grouped}
\resizebox{\textwidth}{!}{%
\begin{tabular}{ll|rrrrrr|rrrrrr|rrrrrr|}
\cline{3-20}
 &
   &
  \multicolumn{6}{c|}{Baseline} &
  \multicolumn{6}{c|}{Monitoring High} &
  \multicolumn{6}{c|}{Tracing High} \\ \cline{3-20}
 &
   &
  \multicolumn{1}{c|}{Primary} &
  \multicolumn{5}{c|}{Auxiliary} &
  \multicolumn{1}{c|}{Primary} &
  \multicolumn{5}{c|}{Auxiliary} &
  \multicolumn{1}{c|}{Primary} &
  \multicolumn{5}{c|}{Auxiliary} \\ \cline{3-20} 
 &
   &
  \multicolumn{1}{c|}{\rotatebox{90}{All}} &
  \rotatebox{90}{Elastic.} &
  \rotatebox{90}{Jaeger} &
  \rotatebox{90}{OTel} &
  \rotatebox{90}{Prom.} &
  \rotatebox{90}{Other} &
  \multicolumn{1}{c|}{\rotatebox{90}{All}} &
  \rotatebox{90}{Elastic.} &
  \rotatebox{90}{Jaeger} &
  \rotatebox{90}{OTel} &
  \rotatebox{90}{Prom.} &
  \rotatebox{90}{Other} &
  \multicolumn{1}{c|}{\rotatebox{90}{All}} &
  \rotatebox{90}{Elastic.} &
  \rotatebox{90}{Jaeger} &
  \rotatebox{90}{OTel} &
  \rotatebox{90}{Prom.} &
  \rotatebox{90}{Other} \\ \hline
\textit{$E_{\mathrm{compute}}$} &
  [Wh] &
  \multicolumn{1}{r|}{58.2} &
  0.9 &
  0.4 &
  4.3 &
  0.3 &
  5.1 &
  \multicolumn{1}{r|}{56.8} &
  0.8 &
  0.3 &
  4.3 &
  1.7 &
  5 &
  \multicolumn{1}{r|}{54.4} &
  12.9 &
  4.8 &
  4.4 &
  0.3 &
  4.7 \\
$E_{\mathrm{network}}$ &
  [Wh] &
  \multicolumn{1}{r|}{8.5} &
  0.6 &
  0.1 &
  10.7 &
  0.2 &
  0.1 &
  \multicolumn{1}{r|}{8.5} &
  0.6 &
  0.1 &
  10.7 &
  2.4 &
  0.1 &
  \multicolumn{1}{r|}{8.4} &
  29.3 &
  0.7 &
  10.7 &
  0.2 &
  0.1 \\
$E_{\mathrm{storage}}$ &
  [Wh] &
  \multicolumn{1}{r|}{-} &
  0.1 &
  - &
  - &
  - &
  4.1 &
  \multicolumn{1}{r|}{-} &
  0.1 &
  - &
  - &
  0.3 &
  4 &
  \multicolumn{1}{r|}{-} &
  2.4 &
  - &
  - &
  - &
  4.1 \\ \hline
$E_{\mathrm{total}}$ &
  [Wh] &
  \multicolumn{1}{r|}{66.7} &
  1.5 &
  0.4 &
  15.0 &
  0.6 &
  9.2 &
  \multicolumn{1}{r|}{65.4} &
  1.5 &
  0.4 &
  14.9 &
  4.4 &
  9.1 &
  \multicolumn{1}{r|}{63} &
  44.6 &
  5.6 &
  15.1 &
  0.5 &
  8.8
\end{tabular}

}
\end{table}

\subsection{Discussion and Recommendation}
In summary, our results align with previous results provided by Dinga et al. \cite{dinga_energyoverheadobservability_2023}. 
Although their results were already indicating a correlation of network with energy consumption, they only focused on CPU load and RAM usage.
Our model and measurements clearly show this observation, as can be seen in \cref{tab:energy_consumption_per_service_experiment_grouped}, where a high frequency in tracing clearly shows up in the network costs.

Moreover, we also considered storage as a fourth component of energy consumption, so far overlooked when considering the energy cost of services.
While the impact on our SUE was not as serious as for the network, it still added up to 2.3 Wh to the total consumption. 
Hence, an increase that previous approaches measuring only CPU and Memory would have missed, and approaches that considered total consumption would have potentially wrongly attributed.

Our results clearly show that ignoring network and storage implied energy consumption results in an estimation error of up to \SI{41}{\percent} and for auxiliary services this error can even go up to \SI{63}{\percent} in the \texttt{Tracing High} scenario. Taking especially network energy consumption into consideration is crucial when configuring and designing systems.

To compare tracing strategies across workloads and platforms, we advocate reporting \emph{energy per captured trace} (Wh/trace) alongside absolute figures. This normalized KPI helps to avoid misleading conclusions that can stem from divergent load profiles or retention settings.
Moreover, at least for a SUE that is similarly structured as the Open Tememetry Demo, we can clearly see that adding and increasing metrics is not as costly as increasing the tracing frequency.
Hence, practitioners should be very mindful of when and how often they trace their services at high levels.
Moreover, adding a convenient component such as Istio, to reduce the need for manual instrumentation or configuring services for tracing and measurement, also comes at a high cost in terms of energy.
Here, practitioners should carefully evaluate if the time that is saved by such approaches justifies the gained automation and other properties of a service mesh, and consider such approaches as a stopgap measure instead of a permanent solution.
Lastly, our experiments showcase that especially the configuration of auxiliary services can significantly impact the total energy consumption of an application. 
Hence, it is important to consider these often shared platform services when considering sustainable service engineering.

Although we acknowledge that estimating network and storage energy consumption by using energy intensities introduces some inaccuracy, omitting these components entirely trades one uncertainty for another: excluding them avoids potential double counting but guarantees underestimation. We favor explicit, literature-grounded estimates with stated assumptions over silent omission.

\section{Conclusion}\label{ch:conclusion}
This work presents a fine-grained, service-level energy consumption model tailored for microservice-based systems, and integrates it into GOXN, a cloud-native experimentation engine for sustainability evaluation. Through this integration, we provide a practical and extensible framework that enables precise measurement and analysis of energy consumption across microservice applications, allowing to make evidence-based, informed architectural and configuration decisions with respect to energy efficiency.

Our study challenges the prevailing reliance on container-level compute metrics for energy efficiency, namely CPU and memory usage. We demonstrate that such metrics alone can significantly underestimate the energy cost of architectural choices, particularly in auxiliary services, where energy consumption is often dominated by cross-container communication and storage operations. Using a representative cloud-native demo application, we experimentally validate GOXN’s ability to reveal and compare the energy implications of alternative service designs. Notably, our results show that service-level configurations can have a substantial impact on overall energy usage. For observability services in particular, we observed that the dominant contributors to energy overhead stem from network and storage operations rather than CPU or memory usage, leading to discrepancies of up to 63\% when compared to state-of-the-art compute-based estimation methods.

Furthermore, our findings reinforce insights from prior studies regarding observability-related energy costs. Specifically, while increased tracing rates significantly elevate energy consumption, a tenfold increase in monitoring granularity does not produce a comparable effect. Unlike previous approaches, however, our model allows for fine-grained attribution of energy consumption to individual service components, thereby identifying the persistence layer’s network interactions as a critical factor in observed energy variations.

For future work, we aim to broaden our evaluation scope to incorporate other SUEs, such as the one proposed by Dinga et al. \cite{dinga_energyoverheadobservability_2023}, and to enhance the accuracy of our model through targeted experiments focused on the characterization of network and storage energy intensities. In line with Hossfeld et al. \cite{hossfeld_energy_intensity_used_wrong_2024}, we recognize that energy intensity should be modeled as a dynamic function of data volume, rather than as a static constant. Adopting this approach holds promise for improving the precision of indirect energy estimations. Additionally, future research should explore mechanisms for attributing energy consumption along service call chains, potentially through techniques such as network packet tagging or other forms of request-level tracing, to enable per-request energy accounting at a granular level.

\subsubsection{Disclosure of Interests}
The authors have no competing interests to declare that are relevant to the content of this article.

\end{document}